\documentclass[journal=jacsat,manuscript=article]{achemso}
\setkeys{acs}{maxauthors=10, etalmode=truncate}

\usepackage[version=3]{mhchem} 
\usepackage{siunitx}
\usepackage{cleveref}
\usepackage{amsmath}     
\usepackage{amssymb}     
\usepackage{xcolor}
\author{H. Tun\c{c} \c{C}ift\c{c}i}
\affiliation{
    Advanced Research Center for Nanolithography (ARCNL), Science Park 106, 1098 XG Amsterdam, The Netherlands
}

\author{Jonathon Cottom}
\affiliation{
    Advanced Research Center for Nanolithography (ARCNL), Science Park 106, 1098 XG Amsterdam, The Netherlands
}
\alsoaffiliation{
    Institute for Theoretical Physics, Institute of Physics, University of Amsterdam, Science Park 904, 1090 GL Amsterdam, The Netherlands
}

\author{Rachid Hahury}
\affiliation{
    Advanced Research Center for Nanolithography (ARCNL), Science Park 106, 1098 XG Amsterdam, The Netherlands
}

\author{Emilia Olsson}
\affiliation{
    Advanced Research Center for Nanolithography (ARCNL), Science Park 106, 1098 XG Amsterdam, The Netherlands
}
\alsoaffiliation{
    Institute for Theoretical Physics, Institute of Physics, University of Amsterdam, Science Park 904, 1090 GL Amsterdam, The Netherlands
}

\author{Bart Weber}
\email{b.weber@arcnl.nl}
\affiliation{
    Advanced Research Center for Nanolithography (ARCNL), Science Park 106, 1098 XG Amsterdam, The Netherlands
}
\alsoaffiliation{
    Van der Waals-Zeeman Institute, Institute of Physics, University of Amsterdam, Science Park 904, 1098 XH Amsterdam, The Netherlands
}

\title[Adhesion Control through Electric Field-Induced Water Adsorption]{Adhesion Control through Electric Field-Induced Water Adsorption at Oxidized Silicon Interfaces}

\begin{document}

\maketitle

\begin{abstract}

Adhesion plays a pivotal role in computer chip manufacturing, directly affecting the precision and durability of positioning components such as wafer stages. Electrical biasing is widely employed to eliminate floating potential and to enable electrostatic clamping. However, upon electrical grounding adhesion can persist and there is limited knowledge about the nature of this adhesion hysteresis. Here, we investigate potential causes underlying electric field-induced adhesion hysteresis at the interface between an \textit{n}-type AFM tip and a \textit{p}-type silicon sample using atomic force microscopy. Our findings reveal that neither charge trapping nor siloxane bond formation significantly impacts the measured adhesion. Surprisingly, we show that adhesion can be tuned through electric field-induced water adsorption under low relative humidity (RH $< 10\%$). Our results provide new insights into adhesion hysteresis and opportunities for adhesion control.

\end{abstract}

\newpage


\section{Introduction}

As micro- and nanoelectronic systems push toward ever-smaller feature sizes, the semiconductor industry faces increasingly stringent demands on positioning precision.\cite{Fischer2015,Yang2019,He2021,Poiesz2020} In such high-precision positioning, adhesion, friction, and slip between silicon wafers and wafer stages play an ever more critical role. Locally varying adhesion between nano-asperities that touch or nearly touch\cite{Carpick1997,Persson2018,Gnecco2001} can exacerbate friction and wear,\cite{Peng2022,hsia2022,Hsia2020} challenging both the positioning accuracy and the longevity of key components.

Adhesion at nanoscale interfaces\cite{Thimons2021} arises through multiple coupled physical and chemical mechanisms. Chief among these are (i)~short-range interactions, including covalent and hydrogen bonding;\cite{Batteas1999,Batteas2003,Diez-Perez2004,doi:10.1021/acs.langmuir.3c02870,Peng2022,Peng2023} (ii)~water-mediated capillary forces;\cite{Lee2015,Cassin2023,Peng2022,hsia2022} and (iii)~electrostatic interactions.\cite{Gavoille2003,Wang2020,Persson2018} The dominant contribution depends on environmental and material parameters: low humidity conditions favour Coulombic, van der Waals\cite{DelRio2005} and direct chemical bonding\cite{doi:10.1021/acs.langmuir.3c02870}, while moderate to high humidity introduces H-bonding and capillary condensation\cite{Peng2022} that can substantially enhance adhesion. Electrostatic forces become significant when surface charge imbalances or externally applied potentials are present. Parameters such as surface roughness\cite{Thimons2021,Hsia2021}, functionalization\cite{Peng2023}, doping, and ambient humidity\cite{Peng2022} modulate each of these interactions, often in non-additive ways.

The qualitative fingerprints of individual force channels, including dispersion, hydrogen bonding, capillary condensation, and siloxane bridge formation, have been mapped in numerous atomic force microscopy (AFM) and surface force apparatus (SFA) studies.\cite{Rodriguez2012,Crooks1995,Sacha2007,Schwaderer2008,doi:10.1021/acs.langmuir.3c02870,Peng2023,Lee2015} Yet, the reported pull-off forces and interfacial energies vary by orders of magnitude, even for ostensibly identical oxide surfaces. The observed range reflects the extreme sensitivity of each interaction to tip radius, surface roughness, local water concentration, and prior electrical, chemical or mechanical history.  Direct numerical comparison therefore risks obscuring, rather than clarifying, the underlying physics. An important challenge is to develop experimental and theoretical frameworks that can \emph{isolate} each channel while accounting for their inevitable coupling in real time. This open challenge is nicely highlighted by recent rough surface adhesion measurements and modeling\cite{Thimons2021}. While detailed experimentation and analysis enabled the identification of the strength and range of alumina-on-diamond adhesion, the underlying adhesion mechanism and possible importance of electrostatic interactions remain unclear.

Electrostatic contributions thus magnify the complexity of adhesion. Modest fields reorder interfacial water dipoles and alter capillary morphology, whereas high fields can induce soft dielectric breakdown, generate charged defects, and promote irreversible chemical bonding.\cite{Peng2022,Ranjan2023}  The resulting adhesion can relax within milliseconds or persist for hours, depending on the metastability of the induced states.

For many applications, a purely static view of adhesion fails to capture the dynamic and interdependent nature of these interactions. Under operating conditions, surfaces may experience mechanical loading or be subject to active bias, producing time-dependent strain and electric fields that dynamically modulate adhesion. Applied bias can induce strong electrostatic clamping—exploited in MEMS actuation—while simultaneously reconfiguring surface dipoles and interfacial water structure. The resulting adhesion reflects a balance between chemical bonding, capillary forces, and electrostatics, each evolving with the local field and environment. Engineering tribosurfaces with programmable adhesion or friction therefore requires isolating these field- and humidity-dependent contributions under controlled conditions.

Since the invention of the transistor, silicon has remained the material of choice in microelectronics. Its stable native oxide and well-controlled doping enable the formation of high-quality \mbox{Si$|$SiO\textsubscript{2}} interfaces that underpin device performance.\cite{Deal2013,Helms1994,Fussel1996,Zhuravlev2000} Field-driven processes at this interface—including carrier accumulation, band bending, and defect formation—critically influence both electronic and mechanical behavior. In tribological contexts, these same processes can modulate adhesion and friction. For instance, bias-induced depassivation of \emph{P$_\mathrm{b}$} centers alters local charge density, enhancing chemical adhesion via Si–O–Si bridge formation and modifying long-range electrostatic forces.\cite{Ciurea2008,Stacey2019,Lu2016} At the same time, electric fields can reorient adsorbed water molecules, shift the hydrogen-bonding network, or induce local condensation, all of which alter the force landscape.\cite{Lee2015,Cassin2023} Even subtle changes in surface hydration,\cite{Kweon2015} partial siloxane formation,\cite{Peng2023} or transient defect states can measurably impact nanoscale adhesion. Yet the interplay between bias-driven structural modifications—such as \emph{P$_\mathrm{b}$}-center activation or charge trapping—and the dynamic response of interfacial water remains poorly understood. In particular, it is unclear which mechanism governs the persistence of elevated adhesion after bias removal. While siloxane bridging and trap-assisted electrostatics are plausible candidates, field-induced restructuring of the interfacial water network could produce similar hysteresis signatures. Each channel is associated with a distinct interaction strength, range, and relaxation time, offering a potential route to disentangle their relative contributions.

In this work, we systematically investigate how applied bias modulates adhesion at the self-mated \mbox{Si$|$SiO$_\mathrm{2}$} interface. Using AFM under controlled humidity, we measure sub-nanonewton adhesion force changes as a function of applied voltage and time. This allows us to quantify both the magnitude and persistence of adhesion hysteresis under varying environmental conditions. We find that the elevated adhesion persists long after bias removal, with a pronounced dependence on relative humidity (RH). The combined magnitude, relaxation kinetics, and RH-dependence strongly support field-induced restructuring of interfacial water as the primary driver of adhesion memory. These findings highlight how electric fields dynamically reconfigure interfacial chemistry and water structuring in silicon-based systems, providing a quantitative foundation for engineering field-programmable tribosurfaces.


\section{Experiment}

\begin{figure}[!t]
\begin{center}
\includegraphics[scale=0.6]{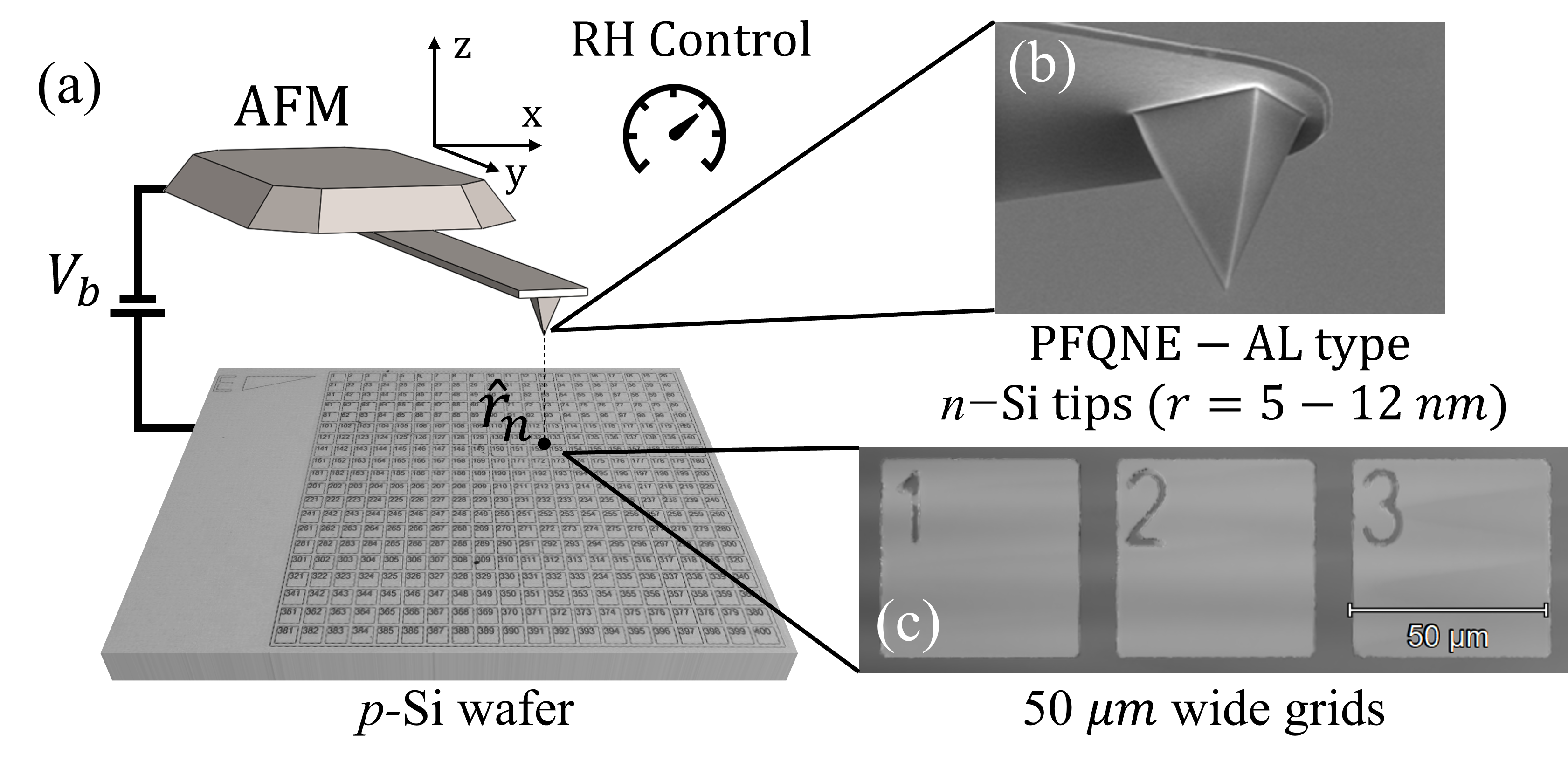}
\caption{(a) Simple schematic for the history-dependent adhesive force measurements using atomic force microscopy (AFM) on p-type silicon (\textit{p}-Si) wafers using (b) Bruker PFQNE-AL type probes carrying n-type antimony doped Si (\textit{n}-Si) tips with a radius of 5-12 nm\cite{bruker_pfqne_al}. The setup allows setting the relative humidity (RH) stable at a chosen level in the chamber. Bias voltages $V_b$ in the rage of 0-10V are applied to the sample prior to adhesion measurements. (c) 50 $\mu$m wide grids etched on the \textit{p}-Si are for precise tip positioning on the surface.}
\label{fig:afm-scheme}
\end{center}
\end{figure}

In order to focus on adhesion mechanisms, rather than the interplay between adhesion mechanisms and surface topography, we performed adhesion measurements at single-asperity interfaces under controlled conditions. We used the Icon Dimension model Bruker AFM system (Figure \ref{fig:afm-scheme}a) with Bruker PFQNE-AL type probes carrying n-type antimony doped Si (\textit{n}-Si) tips with a radius of 5-12 nm (Figure \ref{fig:afm-scheme}b). The tip is mounted on a triangular cantilever with a spring constant of 0.8 N/m and dimensions of approximately 42 $\mu$m in length and 40 $\mu$m in base width.

We collected data from topography and adhesion channels in the PeakForce mode of the Bruker Icon Dimension system, which is placed in a chamber wherein we kept the relative humidity stable in the range of 2\%--80\% using the proUmid Modular Humidity Generator (MHG) 100 system. We used naturally oxidized p-type silicon samples, \textit{p}-Si (Siegert Wafer, $<$100$>$ orientation, boron doped, single side polished, 500-525 $\mu$m thick, 1-10 $\Omega$ resistivity) with 50 $\mu$m wide etched grids (Figure \ref{fig:afm-scheme}c). The naturally oxidized silicon tip and sample surfaces enabled us to investigate the open question of how electric fields influence adhesion in a system for which effects such as charge trapping and water absorption are well understood.

The adhesion force ($F_a$) measurement through PeakForce mode represents the absolute values of the minimum measured force value ($F_{min}$), as given in Eqn.(\ref{eq:adhesion}) below.

\begin{equation}
\label{eq:adhesion}
F_a(\hat{r}_n)= \left| F_{\text{min}}(\hat{r}_n) \right|
\end{equation}

\begin{figure}[!t]
\begin{center}
\includegraphics[scale=0.485]{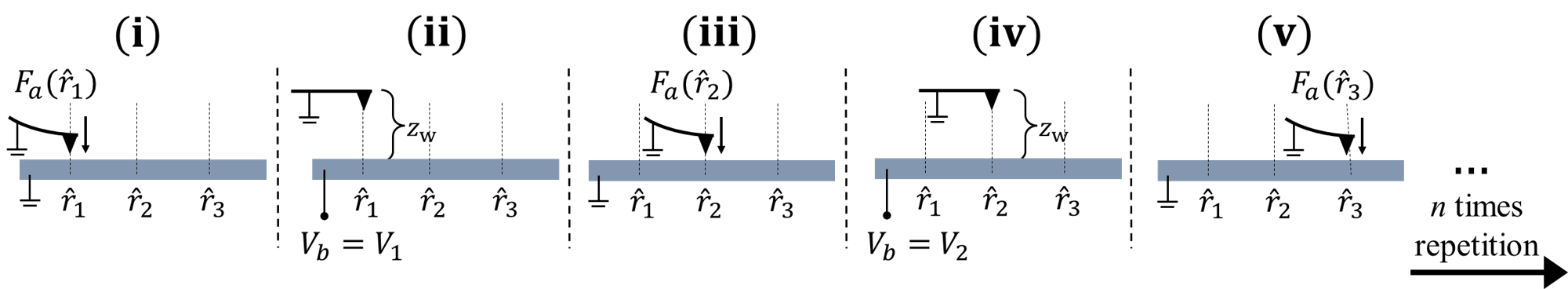}
\caption{Protocol for adhesion hysteresis measurements. (i) Adhesion measurement ($F_a({\hat{r}_1})$) conducted at $\hat{r}_1$ while the tip and the sample are grounded. (ii) Withdrawal of the tip to $z_w = 1$ mm separation at $\hat{r}_1$, after which $V_b = V_1$ is applied for 30 s. (iii) Adhesion measurement ($F_a({\hat{r}_2})$) conducted at $\hat{r}_2$  while the tip and the sample are grounded. (iv) Withdrawal of the tip to $z_w = 1$ mm separation at $\hat{r}_2$, after which $V_b = V_2$ is applied for 30 s. (v) Adhesion measurement ($F_a({\hat{r}_3})$) conducted at $\hat{r}_3$  while the tip and the sample are grounded. The steps (i-v) are repeated for a desired number of measurements. The distance between the positions where the $V_b$ ($\hat{r}_{n-1}$) is applied and $F_a$ ($\hat{r}_n$) measured is about 50 $\mu$m ($\Delta {r} = |\hat{r}_n - \hat{r}_{n-1}| \approx$ 50 $\mu$m).}
\label{fig:protocol}
\end{center}
\end{figure}

To systematically study how electric fields can influence adhesion as a function of time, we developed an adhesion hysteresis measurement protocol that involves several steps as detailed in Figure \ref{fig:protocol}. We first grounded the tip while the wafer is biased, thereby generating an electric field between tip and wafer for 30 seconds. After applying this bias pulse, we engaged the tip to the surface for an adhesion measurement. We repeated this process \( n \) times at different grids \(\hat{r}_n\) (Figure \ref{fig:afm-scheme}c), moving the tip to a neighboring grid for each adhesion force measurement \( F_a(\hat{r}_n) \) while varying the previously applied bias voltage \( V_b(\hat{r}_{n-1}) \). We conducted each adhesion measurement on a new area on the wafer to avoid an influence of scanning induced changes to the wafer on the adhesion measurements. 

When the sample is biased, an electric field is created between the biased wafer and the grounded AFM tip. The AFM tip, including its cantilever and chip, is a millimetric object, while the gap between the tip and wafer is approximately 1 mm. We therefore expect the electric field to cover a millimetric area with a strength of a few volts per millimeter, or a few thousand volts per meter (\( E = \frac{V}{d} \), where \( V = 10 \, \text{V} \) and \( d = 1 \, \text{mm} \)).

\section{Results}

\subsection{Influence of Applied Bias on Adhesion}

\begin{figure}[!t]
\begin{center}
\includegraphics[scale=0.525]{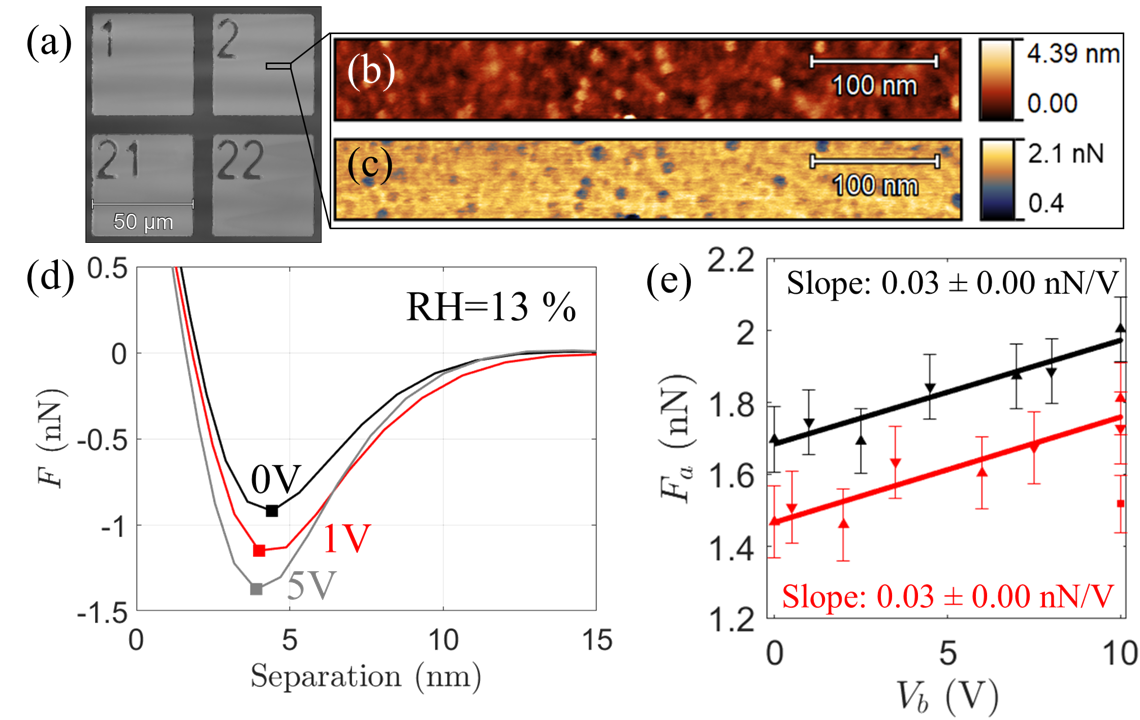}
\caption{
(a) Optical micrograph showing the grid locations where (b) topography (root mean square roughness: 1.13\,nm\,±\,0.2\,nm) and (c) adhesion force ($F_a$) maps (average $F_a = 1.57$\,nN\,±\,0.08\,nN) were acquired over a $512\,\text{nm} \times 32\,\text{nm}$ area with $256 \times 16$ pixel resolution. 
(d) Force--distance curves measured between an \textit{n}-Si AFM tip and a \textit{p}-Si wafer at distinct surface locations $\hat{r}_n$, each corresponding to a single pixel. The black curve shows the initial measurement at $\hat{r}_1$ (before biasing), while the red and gray curves correspond to measurements at $\hat{r}_2$ and $\hat{r}_3$ after applying $V_b = 1\,\text{V}$ and $5\,\text{V}$, respectively, for 30\,s. 
(e) Averaged adhesion force values ($F_a$) from multiple scan areas across different grids, plotted as a function of applied bias $V_b$. Upward triangles represent the forward sweep (0--10\,V), and downward triangles the reverse sweep (10--0\,V). An independent repetition (shown in red) demonstrates that after a 90\,min rest following the 10\,V measurement, $F_a$ decreases (square), but recovers upon reapplying 10\,V (rightmost red triangle pointing down), indicating reversible hysteresis. 
All measurements were performed at 13\% relative humidity with the sample grounded; $V_b$ was applied only between measurements. Fresh \textit{n}-Si AFM tips (Bruker PFQNE-AL) were used for each experiment. The error bars reflect the standard deviation in adhesion force across the measured areas, which is largely caused by variations in surface topography as illustrated by the correlation between adhesion and topography in (b) and (c). 
}

\label{fig:experiment1}
\end{center}
\end{figure}

Figure~\ref{fig:experiment1}a--c provide an overview of the results obtained by applying the protocol described above to the interface between the n-Si tip and the p-Si sample. The optical micrograph in (a) shows the grid positions used for data acquisition, while (b) and (c) display representative topography and adhesion force ($F_a$) maps, respectively, acquired over a 512 nm × 32 nm scan area. The root mean square surface roughness within the measurement area was 1.13 nm ± 0.2 nm, and the average adhesion force was $F_a = 1.57$ nN ± 0.08 nN. Local variations in $F_a$ across the scan area ranged from 1.49 to 1.65 nN, closely correlating with surface features visible in the topography map, indicating that nanoscale surface morphology influences the measured adhesion force.

Upon application of the adhesion hysteresis protocol outlined in Figure~\ref{fig:protocol},  we observed a systematic increase in adhesion at progressively higher bias voltages (Figure~\ref{fig:experiment1}d), even though other measurement parameters (normal load, approach speed, environment) remained unchanged. Note that each curve in Figure~\ref{fig:experiment1}d represents a force measurement at a single pixel. To test whether tip wear was responsible for the observed increase in adhesion with bias, we used spatially resolved measurements of averaged adhesion forces, $F_a$, across areas of $\SI{512}{nm} \times \SI{32}{nm}$ (Figure~\ref{fig:experiment1}e). The bias voltage $V_b$ was first incremented in 1\,V steps from 0\,V to 10\,V (black and red upward triangles) and then decremented back to 0\,V (black and red downward triangles). Note that, except for the data points at 0\,V and 10\,V in Figure~\ref{fig:experiment1}e, each point is an average of several measurements, grouped by adjacent bias steps for clarity. A pronounced hysteresis emerged: $F_a$ rose with increasing $V_b$ and reverted toward its original value upon decreasing $V_b$. An independent repeat of this experiment (red symbols) reproduced the trend. After a \SI{90}{\minute} rest following the 10\,V measurement, $F_a$ decreased (red square), but reapplying 10\,V restored the adhesion (rightmost red downward triangle). These observations show that the elevated adhesion---though long-lived---is nevertheless reversible, ruling out irreversible tip wear. Each data point in Figure~\ref{fig:experiment1}e represents the average of multiple adhesion force measurements acquired from scan areas similar to Figure~\ref{fig:experiment1}c, across different grid locations.

\subsection{Humidity-Dependent Adhesion}
\begin{figure}[!t]
\begin{center}
\includegraphics[scale=0.4]{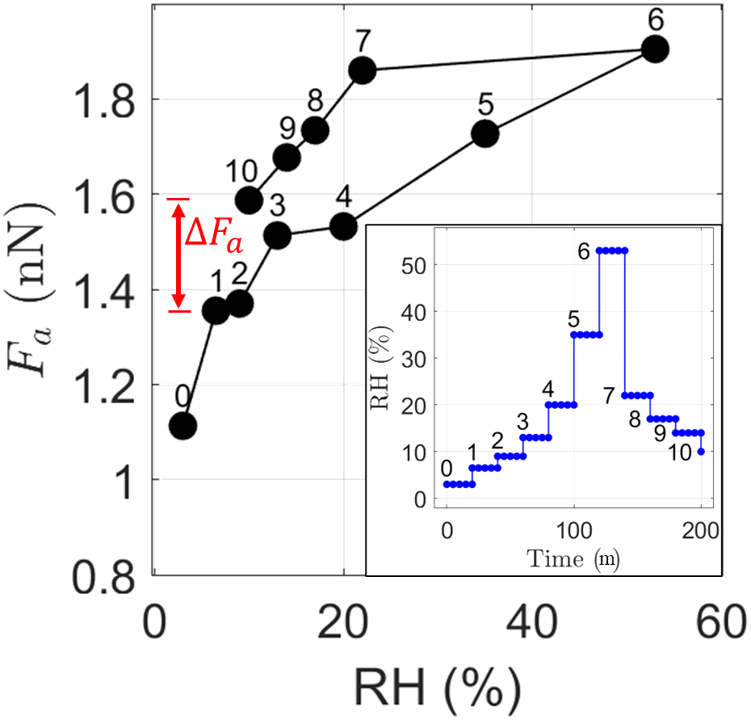 }
    \caption{Adhesion force hysteresis loop versus RH. Eleven measurements (indexed 0--10) were sequentially performed as RH was varied from \SI{2.5}{\percent} to \SI{55}{\percent} and back. The inset tracks the time evolution of RH, each level held for \SI{20}{\minute} prior to measurement. A fresh \textit{n}-Si tip (Bruker PFQNE-AL) and sample were used. The red arrow ($\Delta F_a$) highlights the hysteresis in adhesion attributable to humidity.}
\label{fig:experiment2}
\end{center}
\end{figure}

To investigate if the observed adhesion hysteresis (Figure~\ref{fig:experiment1}) is related to the availability of water in the direct surroundings of the tip/wafer interface, we conducted adhesion measurements while varying the relative humidity. It is well established that water adsorption can strongly affect adhesion at oxide-covered silicon interfaces. Figure~\ref{fig:experiment2} shows $F_a$ measured as RH was first lowered to approximately \SI{2.5}{\percent}, then increased to \SI{55}{\percent}, and finally reduced again. Each humidity step was held for \SI{20}{\minute} to ensure near-equilibration before measurement. Adhesion rose significantly with increasing RH and showed a clear hysteresis upon decreasing RH afterward. Notably, the magnitude of this RH-driven hysteresis ($\sim 250$--\SI{300}{pN}) matches that of the bias-induced hysteresis observed at \SI{13}{\percent} RH (cf.\ Figure~\ref{fig:experiment1}).

\begin{figure}[!t]
\begin{center}
\includegraphics[scale=0.48]{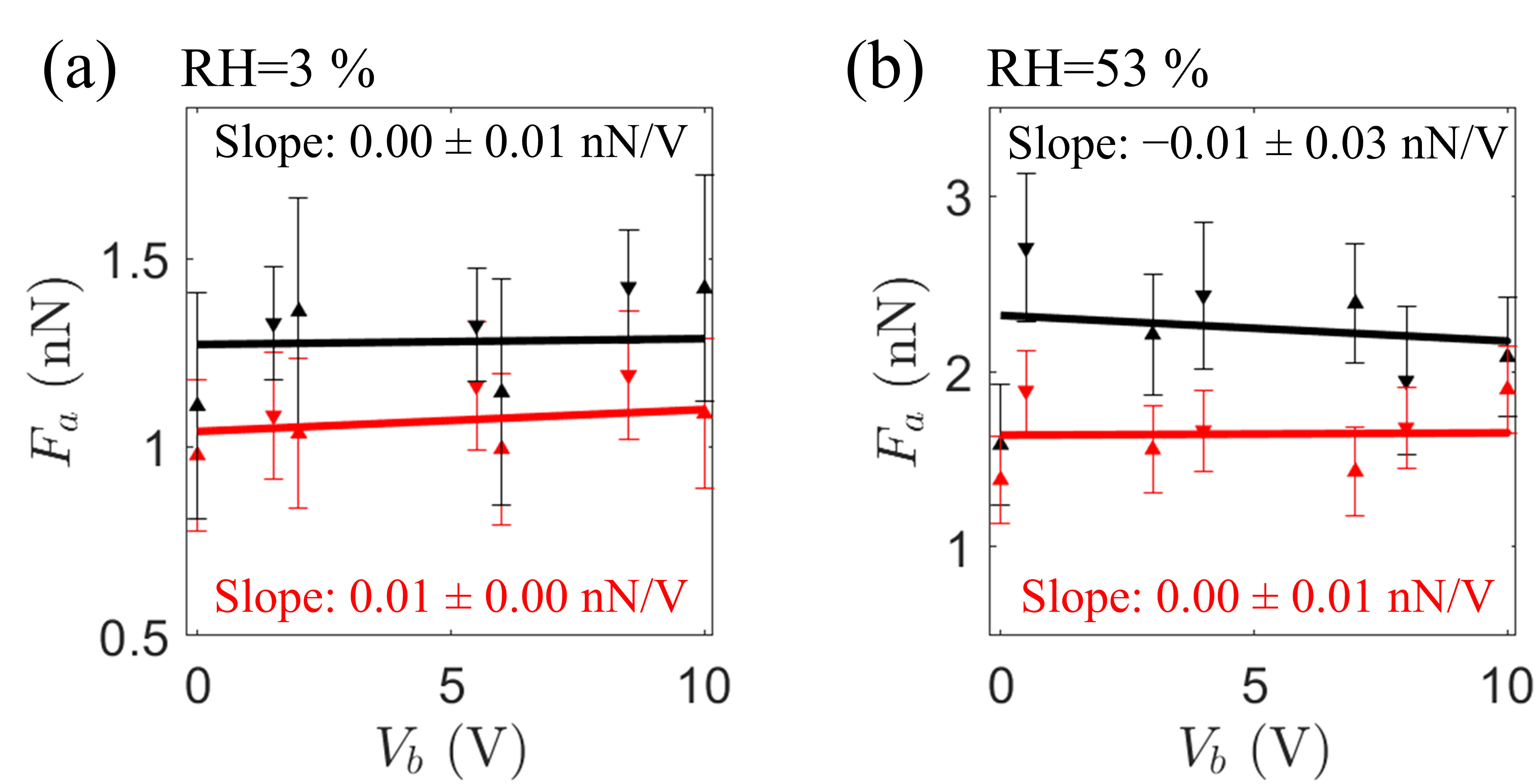}
\caption{(a) and (b) Each data point represents an average of multiple adhesion force measurements ($F_a$) acuired within a $\SI{512}{nm}\times\SI{32}{nm}$ frame, grouped by adjacent bias values (For clarity, the data have been reduced to four representative points per bias sweep by averaging across sequential subsets).  The bias ($V_b$) was increased in steps from 0\,V to 10\,V (upward triangles) and then decreased back to 0\,V (downward triangles). Data in (a) were obtained at \SI{3}{\percent} RH, whereas (b) was performed at \SI{53}{\percent} RH. Both datasets show negligible hysteresis, as opposed to the measurements conducted in at 13\% RH environment (Fig.). The sample was grounded during all measurements, and bias was applied only between measurements. Four fresh \textit{n}-Si tips were used for the experiments.}
\label{fig:experiment3}
\end{center}
\end{figure}

Given the impact of humidity history on adhesion in our system, we explored how the electric field induced adhesion hysteresis depends on relative humidity. To test whether moderate humidity is a prerequisite for bias-induced hysteresis, we performed the same bias-voltage protocol at very low (\SI{3}{\percent}) and relatively high (\SI{53}{\percent}) humidity, shown in Figure~\ref{fig:experiment3}. Under these extreme conditions, the hysteresis essentially vanished, suggesting that additional electric-field-driven water adsorption is most effective at intermediate RH. Near-saturated surfaces (e.g., \SI{53}{\percent} RH) cannot accommodate much more water, whereas at very low RH (\SI{3}{\percent}), insufficient water is available to create the same capillary effects.

\clearpage

\section{Discussion}
As mentioned in the introduction, deciphering which mechanism is responsible for adhesion, including the possible importance of electrostatic interactions is an open challenge. Surprisingly, despite identical conditions during the adhesion measurements reported in Figure~\ref{fig:experiment1}, adhesion increases in each successive measurement, suggesting that the adhesion force is influenced by the electric field applied just before each measurement. The geometry, materials and controlled conditions in our experiments enable a detailed analysis of the various mechanisms that may contribute to the measured adhesion hysteresis. 

\subsubsection{Electrostatic interactions}
The literature suggests that various factors, including crystal orientation, defects, dangling bonds, vacancies, dopants, and localized surface stresses in band gap materials, can contribute to charge trapping \cite{Stacey2019}. In the silicon bulk, structural defects can introduce trap states within the band gap, capturing holes\cite{alkauskas2016}. Vacancies, created by missing atoms, also produce energy states capable of trapping holes, while specific dopants introduce acceptor states within the band gap that act as potential hole traps\cite{hind2025}. At the silicon/oxide interface, certain crystal orientations can increase the density of surface states, enhancing the likelihood of hole trapping\cite{shao2019}. Additionally, dangling bonds at the interface generate localized states within the band gap that are especially prone to trapping holes, particularly when depassivated by applied bias. 

Other factors, such as localized surface stresses near interfaces, can modify the band structure, creating states that promote hole trapping. Together, these influences highlight the complex mechanisms that could lead to adhesion hysteresis through trapped charges in the silicon wafer.

Hole trapping in bulk silicon can lead to the formation of mirror charges (mobile electrons) in the \textit{n}-Si tip, potentially generating electrostatic adhesion. Prior studies, such as those by Falster and Voronkov\cite{Falster2000}, indicate that the density of intrinsic defects of silicon, from $10^{4}$ to $10^{6}~\mathrm{cm}^{-3}$ , is negligible compared to the density of doping-induced alterations found in the bulk of the \textit{p}-Si sample, from $1 \times 10^{18}$ to $4 \times 10^{19}~\mathrm{cm}^{-3}$\cite{Lew2004}. Thus, our analysis focuses on defect characteristics specifically induced by boron doping.

Electrostatic interactions between the tip and the substrate should be dominated by charges present close to the tip-on-substrate interface. The short Debye length of doped silicon will cause strong screening of any charges separated by more than a few nanometers from the contact\cite{Teng1985}. Even in the oxide layer on top of the silicon bulk,  only those charges trapped within a distance of order tip radius should matter, as charges further away engage in weaker Coulombic attraction with only a minor component in the normal direction due to the geometry of the interface. If the observed hysteric increase in adhesion force of 250–300 pN results from an additional electrostatic force, we can obtain a rough estimate of how many charge pairs are involved through Coulomb's law. Coulomb’s law (\(|F_e| = k_e \frac{|q_1 q_2|}{\epsilon_r z_e^2}\), where $q$ represents the charges and counter charges (in units Coulomb),  $k_e=9.0 \cdot 10^9$ $Nm^2C^{-2}$ is the Coulomb constant and $\epsilon_r$ is the relative permitivity of the medium separating the charges), suggests that about two elementary charges and counter charges trapped within the interaction volume between the tip and sample could generate the observed force when the charges are separated by a silicon dioxide film (\(z_e \approx 1~\mathrm{nm}\), $\epsilon_r=3.9$). The precise distance and medium across which charges can interact will of course depend on the exact charging scenario; screening of the interaction between tip and sample charges may take place through water films, oxide layers and silicon bulk. 

We next estimate how many trapped charges could be present within the volume of the wafer that can electrostatically interact with the tip. Approximating this volume as a hemisphere with radius 5 nm and multiplying the hemisphere volume with a typical doping density of \(N_A = 1 \times 10^{19} \, \mathrm{cm^{-3}}\) we estimate that, interestingly, about 10 doping sites are available within the wafer to interact with counter charges in the tip; enough to generate Coulombic attraction of order 300 pN. However, a study by Jang \textit{et al.}\cite{Jang2016} states that the lifetime of trapped holes at boron defect sites is in the millisecond range. This lifetime is far too short to account for the hysteresis observed in our adhesion measurements, which extends over minutes to hours. Such timescales align more closely with mechanisms like charge trapping at oxide interfaces\cite{Lee1993} or water desorption\cite{Peng2022}, both of which are known to occur on significantly longer timescales.

According to Lee \textit{et al.} \cite{Lee1993}, electron trapping within the native silicon oxide layer is negligible, whereas hole trapping is significant due to the low mobility of holes in native oxides and limited tunneling annealing. The reduced mobility of holes in native oxides arises from their high effective mass and interactions with trap states, which further enhance hole trapping. Furthermore, the low tunneling rates in native oxides inhibit trapped holes from easily returning to a conductive state, leading to increased hole accumulation.

The typical density of hole traps in a native oxide layer on \textit{p}-Si is approximately \(10^{11} \, \mathrm{cm^{-2}}\) \cite{Lu2014}. This translates to a maximum density of about \(10^{-3} \, \mathrm{nm^{-2}}\) within the tip-sample interaction area. We approximate the area of the silicon oxide/bulk interface with which the tip can interact electrostatically as a disk with radius 5 nm and thus obtain approximately 0.1 charges within the interaction volume. However, this density is an order of magnitude too low to generate the observed 250–300 pN adhesion hysteresis. We thus also consider interfacial siloxane bonds as a potential mechanism contributing to the observed adhesion hysteresis.

\subsubsection{Short-range interactions: Covalent and H-bonding}
Previous studies \cite{Peng2022} suggest that bonding effects\cite{DANG2025205915}, such as siloxane bond formation across silicon interfaces\cite{doi:10.1021/acs.langmuir.3c02870}, occur when two hydroxyl groups on opposing surfaces meet and bond, generating a water molecule in the process. However, such interfacial siloxane bond formation is expected only under conditions involving freshly plasma-cleaned surfaces and minimal relative humidity (RH $<$ 0.5\%). In our experiments, the silicon samples are not plasma-cleaned, and measurements are conducted at significantly higher humidity levels. Furthermore, the force required to break a single siloxane bond is approximately 700~pN\cite{Peng2023,Schwaderer2008}. In contrast, the adhesion forces observed in our study exhibit a gradual change within the range of 250–300~pN at RH = 13\%, which is insufficient to break even a single siloxane bond. Thus, the gradual increase in adhesion observed is inconsistent with the bonding and rupture dynamics of siloxane bonds, leading us to conclude that interfacial siloxane bonds do not contribute to adhesion hysteresis under our experimental conditions.

While siloxane bonds do not appear to play a role in our system, it is expected that water films on the silicon surfaces can contribute to the measured adhesion, for example through the formation of hydrogen bond networks\cite{Peng2022, zheng2024}. In this scope, parameters such as relative humidity (RH) \cite{Kweon2015} and pH \cite{Batteas1999, Batteas2003} are important. At the RH levels used in our measurements (Figure \ref{fig:experiment1}), water films of approximately 4 \text{\AA} thickness are expected to form on the naturally oxidized \textit{p}-Si surface\cite{Asay2005}. Water adsorption on silicon is known to be history-dependent, as it takes time for water molecules to desorb once the humidity is lowered\cite{Peng2022}.

\subsubsection{Water capillary interactions}
The results in Figure~\ref{fig:experiment2} reveal an increase in adhesion as the humidity rises. We attribute the increase in adhesion with increasing relative humidity to the formation of capillary bridges and enhanced hydrogen bonding networks at the interface. To illustrate this, we make a rough estimate of the capillary force exerted by the capillary bridge \(F_c\) at the interface between the sharp \textit{n}-type silicon conical AFM tip (\(R_{\text{tip}} = \SI{5}{\nano\meter}\)) and the \textit{p}-type silicon wafer with native oxide using \cite{mate2019}:
\[
F_c = 4 \pi R_{\text{tip}} \gamma,
\]
where \(\gamma\) is the surface tension of water. We thus estimate the capillary force to be \(F_c \sim \SI{5}{\nano\newton}\), which is of the same order as the adhesion forces measured experimentally.

In addition to the increase in adhesion with increasing relative humidity -which can be attributed to water capillary effects- we also observe adhesion hysteresis when the humidity is first increased and subsequently decreased (Figure~\ref{fig:experiment2}). Interestingly, the RH induced hysteresis in adhesion (Figure~\ref{fig:experiment2}) is similar in magnitude to the adhesion hysteresis that is observed when the wafer sample is biased (Figure~\ref{fig:experiment1}), with both measurements showing a change in adhesion force on the order of several hundred \si{\pico\newton}. Furthermore, it is remarkable that our measurements show identical behavior to that observed for the humidity dependence of friction of macroscopic silicon-silicon interfaces (both in terms of the trend and the hysteresis) involving forces eight orders of magnitude larger than those reported here\cite{Peng2022}. 

\subsubsection{Electric-Field-Induced Water Adsorption}
Water is known to enhance adhesion and friction at naturally oxidized silicon interfaces, particularly during the formation of the first water monolayers at low relative humidity, as reflected in the first data points of Figure \ref{fig:experiment2}. At approximately 10\% RH, a monolayer of water is expected to form on the naturally oxidized \textit{p}-Si surface\cite{Asay2005}. The electric field generated during the sample bias experiments (Figure \ref{fig:experiment1}) likely promotes increased water adsorption on the biased wafer\cite{Sacha2007}. Polar water molecules in the air gap between the wafer and the tip are attracted to wafer and tip due to the electric field. While the electric field in our setup (approximately \(E = 10^4 \, \mathrm{V/m}\)) is two to three orders of magnitude lower than in typical electrowetting studies\cite{Mugele2005} (\(10^6 - 10^7 \, \mathrm{V/m}\)), the electric field can nonetheless induce water polarization and adsorption. The resulting increase in water availability at the interface enhances adhesion and likely contributes to the observed adhesion hysteresis.

If electric-field-induced water adsorption drives the observed adhesion hysteresis at \(13\%\) RH, we would expect the hysteresis to diminish at higher humidity levels (\(>\!40\%\)), where adhesion becomes less sensitive to RH or sample biasing due to reduced Laplace pressure differences, or at lower humidity (\(<\!5\%\)), where insufficient environmental water is available for significant adsorption on the silicon surfaces. Our measurements (Figure \ref{fig:experiment3}) indeed indicate no significant adhesion hysteresis at either very low or high humidity, supporting the hypothesis that moderate RH levels are necessary for electric-field-induced water adsorption to influence adhesion.

Furthermore, in support of our interpretation, Figures \ref{fig:experiment2} and \ref{fig:experiment3} reveal that adhesion is generally higher at elevated humidity levels. If charge trapping is of importance for the adhesion in our system, we expect higher adhesion at low humidity, because at low humidity less water is available to screen the electrostatic interaction between the tip and the sample\cite{Kweon2015}. However, we observe the opposite trend, with adhesion increasing as humidity rises, supporting the conclusion that water adsorption, rather than charge trapping, plays a central role in enhancing adhesion.

We thus propose that in our adhesion hysteresis experiments, the electric field induced by the applied bias polarizes water molecules in the gap between tip and sample, promoting their adsorption. The increased presence of water on the silicon surfaces exposed to low humidity environment (\(13\%\) RH) enhances the capillary interaction between the tip and the sample.

\section{Conclusion}

In this study, we explored the hysteric effect of sample-biasing on adhesion forces between an \textit{n}-type AFM tip and a \textit{p}-type silicon sample under varying bias voltages at different humidity levels. Our results reveal a significant adhesion hysteresis influenced by applied bias voltage and environmental conditions. We analyzed the observations highlighting the relative importance of charge trapping, siloxane bonding and capillary adhesion.

Bulk or oxide layer charge trapping does not significantly contribute to the observed adhesion hysteresis. Given the typical timescale of hole trapping and the defect concentration wherein charge carriers can remain in the \textit{p}-type silicon, bulk and oxide electrostatic forces from trapped charges are insufficient to explain the observed adhesion hysteresis.

Previous studies have suggested that siloxane bond formation can lead to significant adhesion across silicon interfaces. However, the force required to break an individual siloxane bond (700 pN) is considerably higher than the magnitude of the observed adhesion hysteresis (250--300 pN). Thus, we conclude that siloxane bond formation does not play a role in our measurements. 

Our results indicate that adhesion hysteresis is caused by electric field induced water adsorption, which results in increased capillary adhesion. This adhesion hysteresis is most pronounced at low relative humidity (around \(10\%\)) and can last for minutes after applying the bias voltage. In future studies, the magnitude of the adhesion hysteresis could be further increased, for example by adjusting the humidity level, applied bias and gap. Furthermore, while charge trapping did not influence the adhesion in our measurements, there may be opportunities to further explore the potential of charge trapping for adhesion control, for example by introducing higher densities of charge traps in the sample, or by increasing the electric field strength in the oxide. Disentangling the various mechanisms that contribute to adhesion is an open challenge which we address here by conducting experiments with controlled geometry, materials and environmental conditions including relative humidity and electric field. A deeper understanding of which adhesion mechanisms dominate under what conditions will lead to new opportunities for adhesion control, which is crucial in the semiconductor industry.

\begin{acknowledgement}

This work was conducted at the Advanced Research Center for Nanolithography, a public-private partnership between the University of Amsterdam (UvA), Vrije Universiteit Amsterdam (VU), Rijksuniversiteit Groningen (RUG), the Netherlands Organization for Scientific Research (NWO), and the semiconductor equipment manufacturer ASML. We would like to gratefully acknowledge Arend-Jan van Calcar for his contributions to the experimental setup, Prof. Dr. Esther Alarcon and Dr. Melanie Micali from AMOLF for their valuable insights into adhesion measurements, and Bob Drent and Laura Juskenaite from AMOLF NanoLab for providing samples. We thank Sergey Lemeshko and Marcel Laarhoven for their assistance with the Icon Dimension model Bruker AFM system. We thank Prof. Dr. Tevis Jacobs and Dr. Félix Cassin from the University of Pittsburgh, Dr. Bart Stel from ASML, Prof. Dr. Rob Carpick from the University of Pennsylvania and Prof. Dr. James Batteas from Texas A\&M for insightful discussions on intermediate results.

\end{acknowledgement}

\newpage
\bibliography{bibliography.bib}

\end{document}